\documentclass{article}

\usepackage{PRIMEarxiv}

\usepackage[utf8]{inputenc} % allow utf-8 input
\usepackage[T1]{fontenc}    % use 8-bit T1 fonts
\usepackage{hyperref}       % hyperlinks
\usepackage{url}            % simple URL typesetting
\usepackage{booktabs}       % professional-quality tables
\usepackage{amsfonts}       % blackboard math symbols
\usepackage{nicefrac}       % compact symbols for 1/2, etc.
\usepackage{microtype}      % microtypography
\usepackage{lipsum}
\usepackage{fancyhdr}       % header
\usepackage{graphicx}       % graphics
\graphicspath{{media/}}     % organize your images and other figures under media/ folder

\usepackage{float}          % enable [H]
\usepackage{placeins}       % enable \FloatBarrier

\title{Highly Detailed and Generalizable Broadleaf Tree Crown Instance Segmentation from UAV Imagery
}

\author{
  \textbf{Mitsutaka Nakada$^{1}$, Takahiko Ikebata$^{1}$, Kengo Ikebata$^{1}$, 
  Yuji Mizuno$^{2}$, Yusuke Onoda$^{3}$,} \\
  \textbf{Ryuichi Takeshige$^{3,4}$, Kyaw Kyaw Htoo$^{3}$, 
  Kanehiro Kitayama$^{3,5}$,} \\
  \textbf{Robert Ong$^{6}$, Masanori Onishi$^{1,3}$} \\
  \\
  $^{1}$ DeepForest Technologies Co., Ltd., Kyoto 600-8006, Japan \\
  $^{2}$ YM Lab., Osaka 542-0081, Japan \\
  $^{3}$ Graduate School of Agriculture, Kyoto University, Kyoto 606-8502, Japan \\
  $^{4}$ Graduate School of Science, Osaka Metropolitan University, \\
  \hspace*{1em}3-3-138 Sugimoto, Sumiyoshi-ku, Osaka 558-8585, Japan \\
  $^{5}$ Faculty of Tropical Forestry, Universiti Malaysia Sabah, \\
  \hspace*{1em}Kota Kinabalu, Sabah 88400, Malaysia \\
  $^{6}$ Forest Research Centre, Sabah Forestry Department, \\
  \hspace*{1em}Sandakan, Sabah 90000, Malaysia
}

\begin{document}
\maketitle

\begin{abstract}
We present a highly detailed instance segmentation model for delineating individual tree crowns in natural broadleaf forests using aerial imagery acquired by unmanned aerial vehicles (UAVs). Tree crown delineation in broadleaf forests is more challenging than in other forest types due to diversity of crown shapes and the lack of clearly defined treetops. To address this issue, we developed a deep-learning-based crown segmentation model trained on high-quality annotated crown outlines. We manually delineated 18,507 crown polygons from orthomosaic images collected across seven forests in Japan by skilled annotators, and developed a model based on Mask2Former with multiple backbone architectures. The best model achieved high segmentation performance in structurally complex broadleaf forests using only RGB imagery. This performance was maintained when applied to geographically distinct forests within Japan, as well as to biologically distinct tropical rainforests in Borneo. These results demonstrate that using a large number of high-quality annotated datasets is critical for achieving detailed and generalizable crown segmentation across diverse forest ecosystems. The developed model has been integrated into DF Scanner Pro, a software that supports practical forest monitoring using UAVs, and this implementation is expected to enable a wide range of users to analyze tree-level information in broadleaf forest from UAVs.
\end{abstract}

% keywords can be removed
\keywords{instance segmentation \and deep learning \and UAV imagery\and broadleaf tree \and tree crown delineation}

\section{Introduction}

Understanding forest conditions at the individual-tree level forms the foundation for evaluating the economic value of forests such as timber volume, carbon stocks, and biodiversity, and implementing sustainable forest management \cite{Koch2006, Nevalainen2017}. However, conventional forest inventories based on field surveys require substantial labor and time, making it difficult to acquire information on all individual trees across large areas \cite{Zhou2020}. In recent years, research and practical applications of UAV-based forest measurement have advanced, offering the advantage of covering areas ranging from tens to hundreds of hectares. In addition, it has become increasingly possible to identify tree species from individual crown imagery, while crown area, along with tree height, plays an important role in estimating diameter at breast height (DBH) at the individual tree level from UAVs \cite{Onishi2018, Htoo2025}. Accordingly, there has been growing interest in the automatic extraction of individual tree crowns over extensive areas with high accuracy.

Individual tree crown delineation has traditionally relied on manual visual interpretation or semi-automatic approaches \cite{Steier2025}. In recent years, numerous semi-automatic methods have been proposed using canopy height models (CHMs) derived from LiDAR (Light Detection and Ranging) or UAV photogrammetry. These approaches are primarily based on image segmentation algorithms, including template matching, the watershed method, and region-growing methods \cite{Zhen2015, Dai2018, Brieger2019, Yang2020}. Many of these methods assume that each tree has a single treetop and generate individual tree crowns by identifying local maxima as treetops \cite{Weinstein2020, Yang2020}.

While these approaches are effective in stands with relatively simple crown structures and clearly defined treetops, their application is not straightforward in broadleaf forests, where crown shapes tend to be irregular. In such forests, crown size and shape vary among tree species and individual trees, and distinct treetops may not be clearly identifiable. As a result, algorithms that assume a single local maximum per tree are prone to over-segmentation or under-segmentation.

Recently, deep learning--based instance segmentation methods have been applied to individual tree crown delineation \cite{Zheng2025}. While these studies report high accuracy and demonstrate the effectiveness of such approaches, limitations remain in the quality of the annotation and evaluation data. In many cases, training data are annotated only coarsely with respect to tree crowns, and the annotations are not created by accurately tracing the crowns down to the branch tips. As a result, the predicted outputs often exhibit large gaps between adjacent tree crowns, and relatively smaller trees located between dominant canopy trees are frequently excluded from detection targets \cite{Do2026, Ball2023}. Moreover, progress in generalizable crown instance segmentation has been limited by the scarcity of tree-level annotated datasets with consistently high quality and accuracy collected across diverse forest environments. This limitation is particularly pronounced in broadleaf forests, where crown shapes are diverse and complex, making it difficult to construct large-scale, high-quality annotation datasets.

The objective of this study is to develop a highly detailed and broadly applicable instance segmentation approach for individual tree crowns in broadleaf forests. To this end, we conducted instance segmentation of individual tree crowns in broadleaf forests using high-resolution RGB imagery acquired by UAVs. Specifically, we created a high-quality annotation dataset in which broadleaf crown boundaries were precisely delineated and used it as training data. To ensure diversity in stand conditions and crown morphology, flight data acquired at multiple sites were used to construct a dataset containing over 18,000 crown instances. Based on these data, we developed an instance segmentation model capable of detailed crown delineation in broadleaf forests. The model performance was subsequently evaluated, and we visualized the inference results at sites in Japan and in Borneo.

\section{Materials \& Methods}

\subsection{Study Areas}

\begin{figure}[t]
\centering
\includegraphics[
width=\linewidth,
height=1\textheight,
keepaspectratio
]{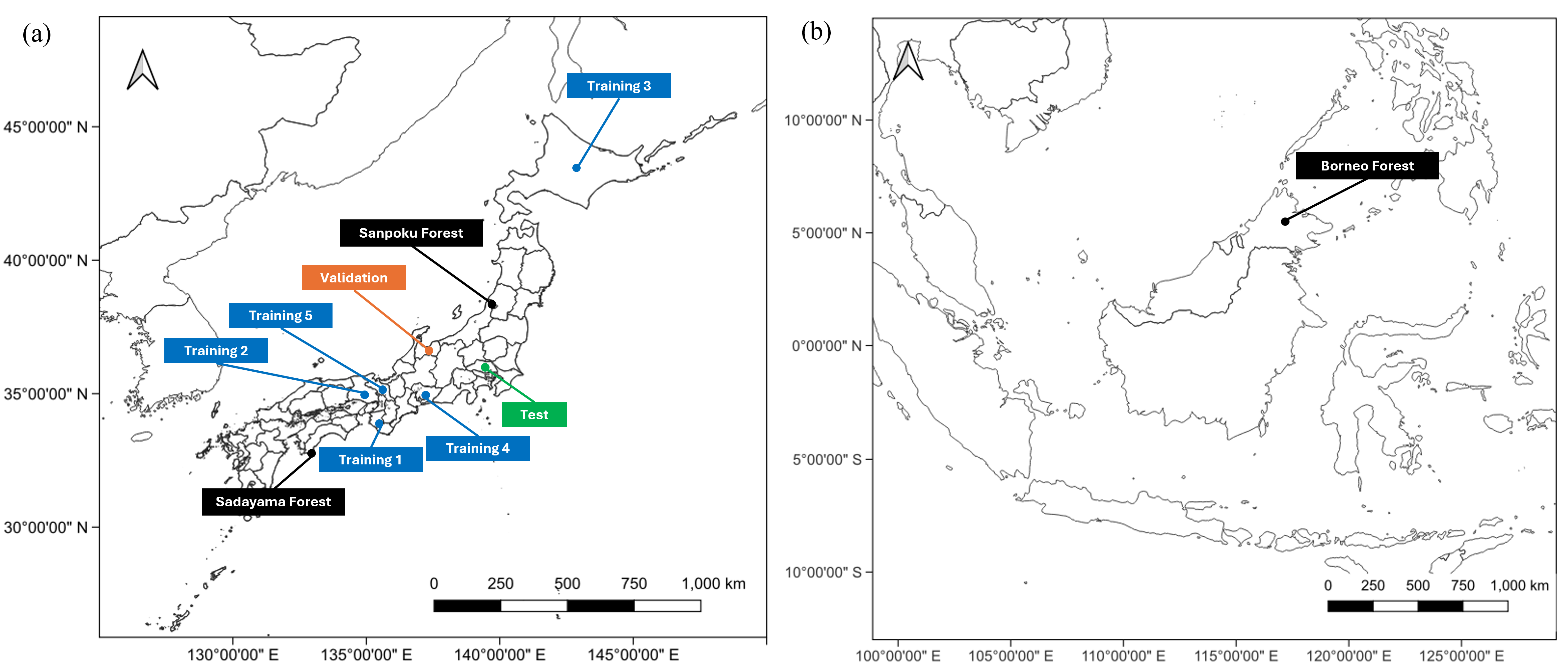}
\caption{Locations of the sites used for model development (training, validation and test) and inference (Sanpoku Forest, Sadayama Forest, and Borneo Forest). As the detailed locations of the model development datasets cannot be disclosed, a pin has been placed at the center of the prefecture.}
\label{fig:study_area}
\end{figure}

The model development datasets (training, validation and test) were collected from seven locations in Japan, most of which are temperate forests (Figure~\ref{fig:study_area}a). The generalization performance of the developed model during inference was evaluated using data from three sites that were not included in the model development datasets: the Sanpoku area of Murakami City, Niigata Prefecture, Japan (Sanpoku Forest; 38$^\circ$24'46''N, 139$^\circ$36'9''E), the Sadayama region in Kochi Prefecture, Japan (Sadayama Forest; 32$^\circ$44'22''N, 133$^\circ$00'05''E) (Figure~\ref{fig:study_area}a), and the Deramakot Forest Reserve in Sabah, Malaysia (Borneo Forest; 5$^\circ$22'24''N, 117$^\circ$25'42''E) (Figure~\ref{fig:study_area}b). Access license to conduct research in Sabah was granted from Sabah Biodiversity Council (JKM/MBS.1000-2/2 JL.D.15(73)) to KK. The forest types at each site were summarized in Table~\ref{tab:forest_type}.

\begin{table}[htbp]
\centering
\caption{Forest type of study site.}
\label{tab:forest_type}

\begin{tabular}{p{3cm} p{10cm}}
\toprule
Site & Forest Type \\
\midrule

Training 1 &
Temperate Mixed Forest (evergreen broadleaf forest and conifer plantation: \textit{Cryptomeria japonica}, \textit{Chamaecyparis obtusa}) \\

Training 2 &
Temperate Mixed Forest (evergreen broadleaf forest and conifer plantation: \textit{Cryptomeria japonica}, \textit{Chamaecyparis obtusa}) \\

Training 3 &
Subarctic Mixed Forest (deciduous broadleaf forest and conifer plantation: \textit{Larix kaempferi}) \\

Training 4 &
Temperate Broadleaf Forest (deciduous and evergreen broadleaf forest) \\

Training 5 &
Temperate Mixed Forest (deciduous broadleaf forest and conifer plantation: \textit{Chamaecyparis obtusa}) \\

Validation &
Temperate Mixed Forest (deciduous broadleaf forest and conifer plantation: \textit{Cryptomeria japonica}) \\

Test &
Temperate Broadleaf Forest (deciduous and evergreen broadleaf forest) \\

Sanpoku Forest \newline (Inference) &
Temperate Mixed Forest (deciduous broadleaf forest and conifer plantation: \textit{Cryptomeria japonica}) \\

Sadayama Forest \newline (Inference) &
Temperate Broadleaf Forest (evergreen broadleaf forest) \\

Borneo Forest \newline (Inference) &
Tropical Rainforest (lowland dipterocarp forest) \\

\bottomrule
\end{tabular}
\end{table}

\subsection{UAV Flight}

We used several types of UAVs and sensors to acquire aerial imagery. For model development datasets acquisition, aerial images and LiDAR data were collected using a Mavic 2 Pro, a Matrice 350 RTK equipped with either a Zenmuse L1 or L2 sensor, and a Mavic 3E (all manufactured by DJI, Da-Jiang Innovations Science and Technology Co., Ltd., Shenzhen, China). The flight parameters for each UAV are summarized in Table~\ref{tab:flight_parameters}.

For inference, the flight in Sanpoku Forest was conducted on June 24, 2024, using a Mavic 3E under the settings shown in Table~\ref{tab:flight_parameters}. In Sadayama Forest, aerial imagery was acquired on August 22, 2023, using a Matrice 300 RTK equipped with a Zenmuse L1 \cite{Takeshige2025}. The flight overlap was set to 80\% in the forward direction (front overlap) and 80\% in the lateral direction (side overlap). The ground sampling distance (GSD) was 2.2 cm, and the flight speed was 5.0 m/s. In Borneo Forest, aerial imagery was acquired on May 21, 2019, using a Phantom 4 Pro (DJI, Shenzhen, China). The flight overlap was set to 80\% in the forward direction (front overlap) and 80\% in the lateral direction (side overlap). The GSD was 3.3 cm.

\begin{table}[htbp]
\centering
\caption{Flight Parameters of UAV Surveys.}
\label{tab:flight_parameters}

\begin{tabular}{p{3.2cm} p{3.5cm} c c c c}
\toprule
Data Class & UAV & Front Overlap (\%) & Side Overlap (\%) & GSD (cm) & Speed (m/s) \\
\midrule

Model Development &
Mavic 2 Pro &
$\geq$85 & $\geq$75 & 2.5 & 5--10 \\

Model Development &
Matrice 350 + \newline Zenmuse L1/L2 &
$\geq$85 & $\geq$75 & 2.5 & 5--10 \\

Model Development &
Mavic 3E &
$\geq$85 & $\geq$75 & 2.5 & 5--10 \\

Sanpoku Forest \newline (Inference) &
Mavic 3E &
85 & 75 & 2.5 & 5--10 \\

Sadayama Forest \newline (Inference) &
Matrice 300 + \newline Zenmuse L1 &
80 & 80 & 2.2 & 5 \\

Borneo Forest \newline (Inference) &
Phantom 4 Pro &
80 & 80 & 3.3 & 5--10 \\

\bottomrule
\end{tabular}
\end{table}

\subsection{UAV Data Processing}

The acquired images were processed using structure-from-motion software Agisoft Metashape Professional (Agisoft LLC, St. Petersburg, Russia) with the following parameters: ``Accuracy'' was set to ``High'' in the ``Align Photos'' step; ``Depth Maps'' was selected as the data source and ``Quality'' was set to ``High'' in the ``Build Point Cloud'' step; ``Point Cloud'' was used as the data source and ``Quality'' was set to ``High'' in the ``Build DEM'' step; and ``Surface'' was set to ``DEM'' in the ``Build Orthomosaic'' step.

\subsection{Model Development Datasets Annotation}

The boundaries of individual tree crowns on the orthomosaic image were visually inspected and manually delineated by experienced annotators using DF Scanner Pro (DeepForest Technologies Co., Ltd., Kyoto, Japan). We conducted the annotation using a tablet with pixel-level precision, and implemented a double-check system in which the results were reviewed by another person. Annotators primarily referred to the orthomosaic images. When canopy height models (CHMs) were available, they were also used to assist crown delineation.

The CHMs were generated differently depending on the data source: LiDAR-derived CHMs were produced through preprocessing in DJI Terra (DJI, Shenzhen, China) followed by DF LAT (DeepForest Technologies Co., Ltd., Kyoto, Japan), while for digital camera data, CHMs were calculated as the difference between DSMs generated via SfM and the 5m resolution DEM (DEM5A) provided by the Geospatial Information Authority of Japan (\href{https://www.gsi.go.jp/kikakuchousei/kikakuchousei40182.html}{DEM5A webpage}).

\begin{figure}[htbp]
\centering
\includegraphics[
width=\linewidth,
height=0.5\textheight,
keepaspectratio
]{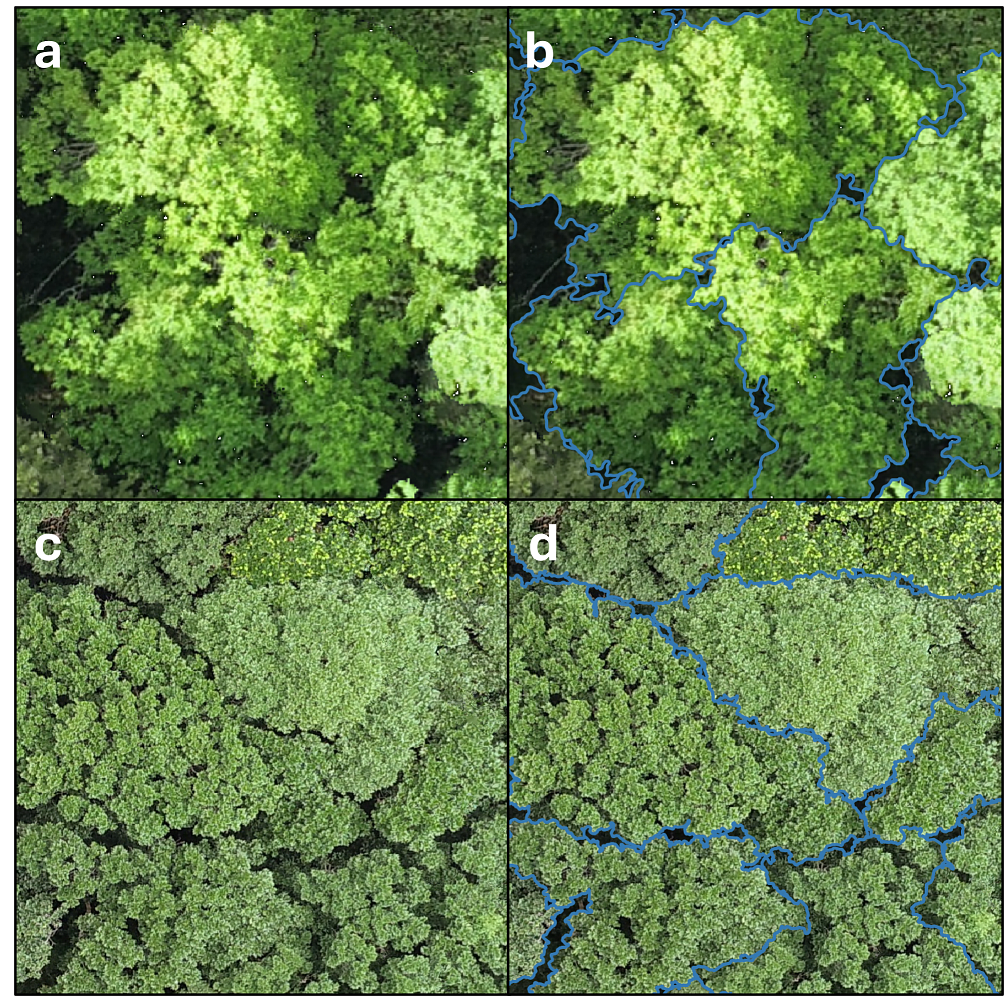}
\caption{Examples of training data annotation for tree crown delineation. (a, c) Original orthomosaic images. (b, d) Corresponding manually delineated tree crown polygons overlaid on the images.}
\label{fig:annotation_example}
\end{figure}

In total, 18,507 crown polygons were prepared as training dataset. An example of the annotated crown polygons is shown in Figure~\ref{fig:annotation_example}.

\subsection{Dataset Construction}

The generated orthomosaic images and their corresponding vector files were converted into the COCO format for model training \cite{Lin2015}. The orthomosaic images were cropped into tiles of 1024 $\times$ 1024 pixels with a 50\% overlap between adjacent tiles. Among the seven model development datasets, five were assigned to the training dataset, one to the validation dataset, and one to the test dataset. The tiled dataset consisted of 3,029 training images, 917 validation images, and 134 test images.

\subsection{Mask2Former}

In this study, Mask2Former was adopted as the instance segmentation architecture \cite{Cheng2022}. Mask2Former is a Transformer-based universal segmentation model that unifies semantic, instance, and panoptic segmentation under a mask classification framework. The model takes multi-scale feature maps extracted from a backbone network as input and employs a query-based Transformer decoder to simultaneously predict instance masks and class labels. Each query corresponds to a potential object, and bipartite matching (Hungarian matching) is applied during training to optimally assign predictions to ground-truth masks. This design enables accurate and stable instance segmentation even in images containing complex backgrounds and densely distributed objects.

\subsection{Training and Evaluation}

In this study, we employed MMDetection v3.3.0, a PyTorch-based object detection framework. The development environment was built on Docker, using a base image that included PyTorch 2.1.0, CUDA 11.8, and cuDNN 8, on which MMDetection was installed. Model training and inference were conducted on an NVIDIA RTX 3090 GPU (24 GB).

Mask2Former was adopted as the instance segmentation model, and pretrained weights provided by MMDetection were used for initialization. Four different backbones were evaluated and compared their segmentation performance: ResNet-50 (R-50), ResNet-101 (R-101), Swin Transformer-Tiny (Swin-T), and Swin Transformer-Small (Swin-S) \cite{He2015, Liu2021}.

The number of classes was set to one (tree crown), and the input image size was 1024 $\times$ 1024 pixels. The batch size was set to 2, and AdamW was used as the optimizer \cite{Loshchilov2019}. The learning rate was set to $1.0 \times 10^{-4}$, and the weight decay to 0.05. The maximum number of training iterations was 368,750, and segmentation mAP on the validation dataset was computed every 5,000 iterations.

The segmentation mAP used in this study follows the COCO evaluation metric and is defined as the mean of the Average Precision (AP) values computed over Intersection over Union (IoU) thresholds ranging from 0.50 to 0.95 in increments of 0.05. In addition, mAP at IoU = 0.50 and IoU = 0.75 (denoted as mAP50 and mAP75, respectively) were also calculated.

Model selection was based on the segmentation mAP on the validation dataset, and the model achieving the highest value was designated as the best model. Finally, this best model was used to compute segmentation mAP, mAP50, and mAP75 on the test dataset for final performance evaluation.

\subsection{Inference}

For inference process, the input orthomosaic images were cropped into tiles of 1024 $\times$ 1024 pixels, with 80\% overlap between adjacent tiles. Each tile was then fed into the trained model for instance segmentation. The confidence score threshold was set to 0.3, and predicted instances with scores below this threshold were excluded.

The overlapping tiles method was applied to reduce boundary artifacts caused by tile-based processing \cite{Payer2019}. Specifically, instances in contact with the outer edges of each tile were removed to suppress the influence of incomplete predictions near tile boundaries.

Furthermore, post-processing was conducted following Chen et al. \cite{Chen2021} to integrate redundant instances detected across adjacent tiles. The Intersection over Union (IoU) was calculated between overlapping instances, and those with an IoU of 0.1 or greater were merged, as they were assumed to represent the same tree crown. The merged instance masks were converted into polygons, and an additional post-processing step was performed to remove overlaps between polygons.

\subsection{Evaluation of Model Generalization}

To evaluate the generalization performance of the trained model, inference was conducted using datasets that were not used in model development process. Specifically, the trained model was applied to orthomosaic images of Sanpoku Forest, Sadayama Forest and Borneo Forest. The inference results were visualized using QGIS 3.34.11, and the results were interpreted by visual inspection.

% \clearpage
\section{Results}

Table~\ref{tab:model_performance} summarizes the segmentation mean Average Precision (mAP) for the single class, ``tree crown,'' across different Mask2Former backbones. Among them, Swin Transformer-Small achieved the highest mAP. This backbone has the largest model size and the highest representational capacity among those evaluated in this study, which likely demonstrates its superior performance. Based on these results, we adopted Swin Transformer-Small as the backbone for all subsequent analyses.

\begin{figure}[H]
\centering
\includegraphics[
width=\linewidth,
height=0.8\textheight,
keepaspectratio
]{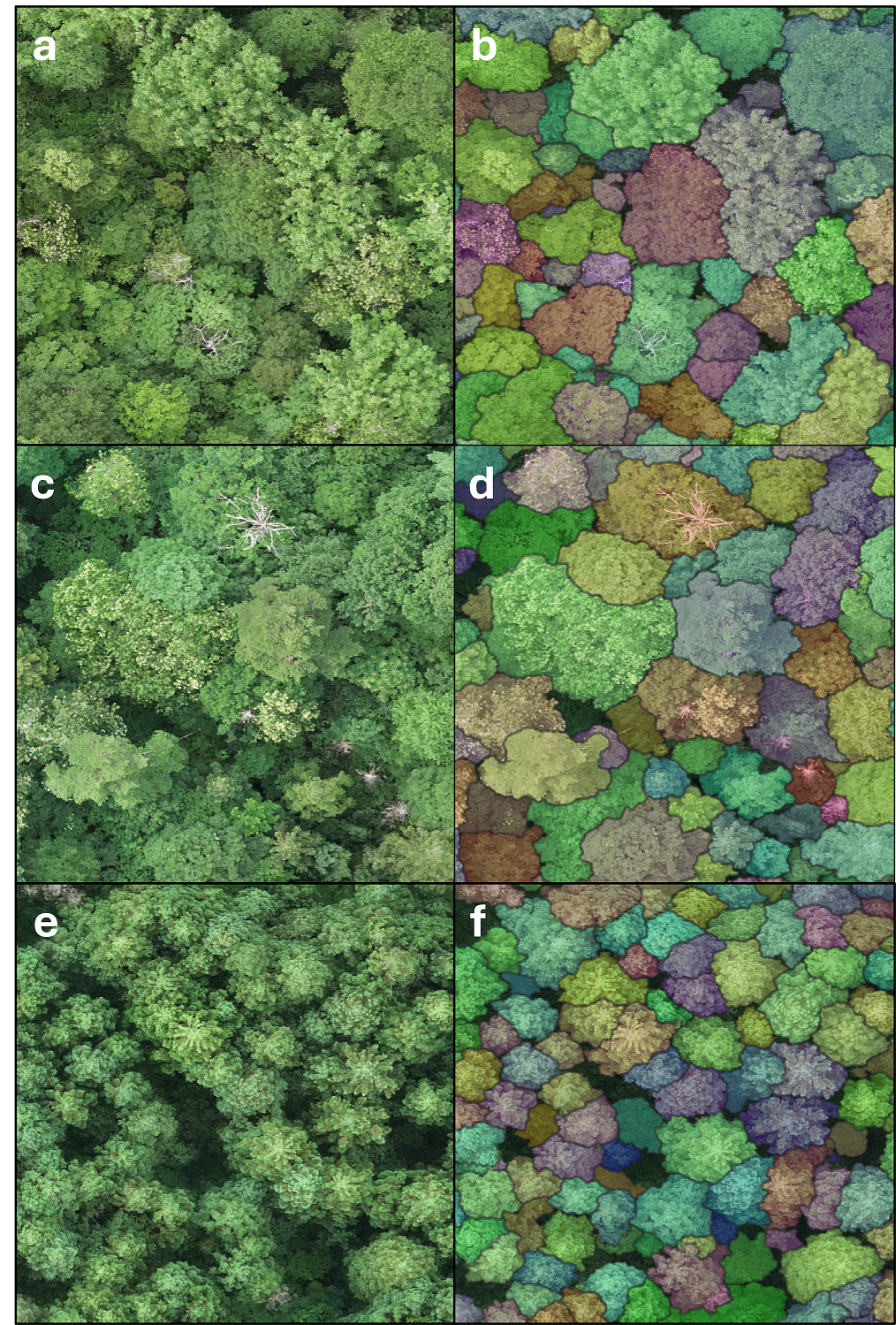}
\caption{Examples of tree crown segmentation results in Sanpoku Forest. (a, c) Original orthomosaic images of broadleaf forest areas. (b, d) Corresponding inference results overlaid on the images shown in (a) and (c), respectively. (e) Original orthomosaic image of a plantation Japanese cedar forest area. (f) Corresponding inference results overlaid on the image shown in (e).}
\label{fig:sanpoku_result}
\end{figure}

\begin{figure}[H]
\centering
\includegraphics[
width=\linewidth,
height=0.8\textheight,
keepaspectratio
]{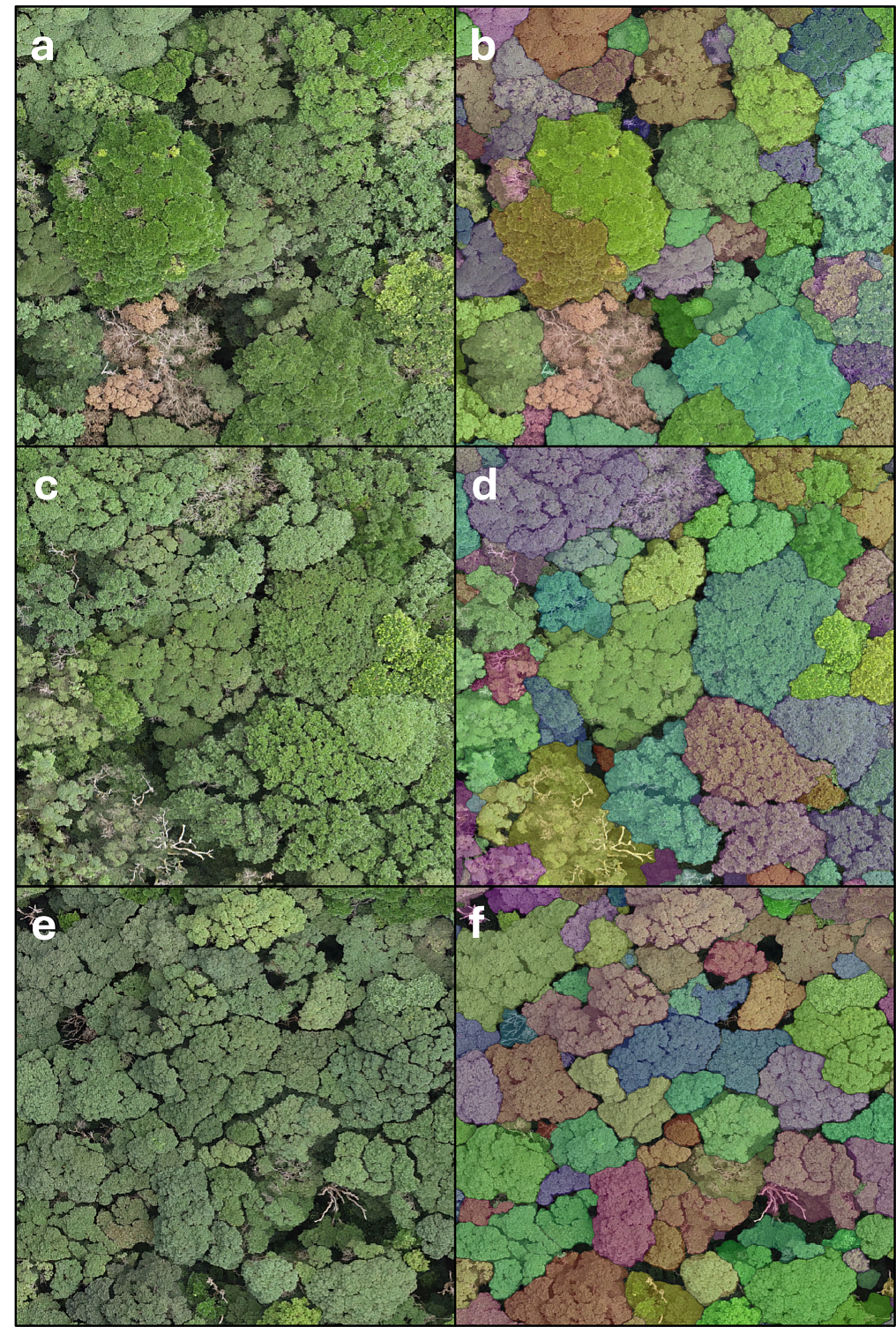}
\caption{Examples of tree crown segmentation results in Sadayama Forest. (a, c, e) Original orthomosaic images. (b, d, f) Corresponding inference results overlaid on the images shown in (a), (c), and (e), respectively.}
\label{fig:sadayama_result}
\end{figure}

\begin{figure}[H]
\centering
\includegraphics[
width=\linewidth,
height=0.8\textheight,
keepaspectratio
]{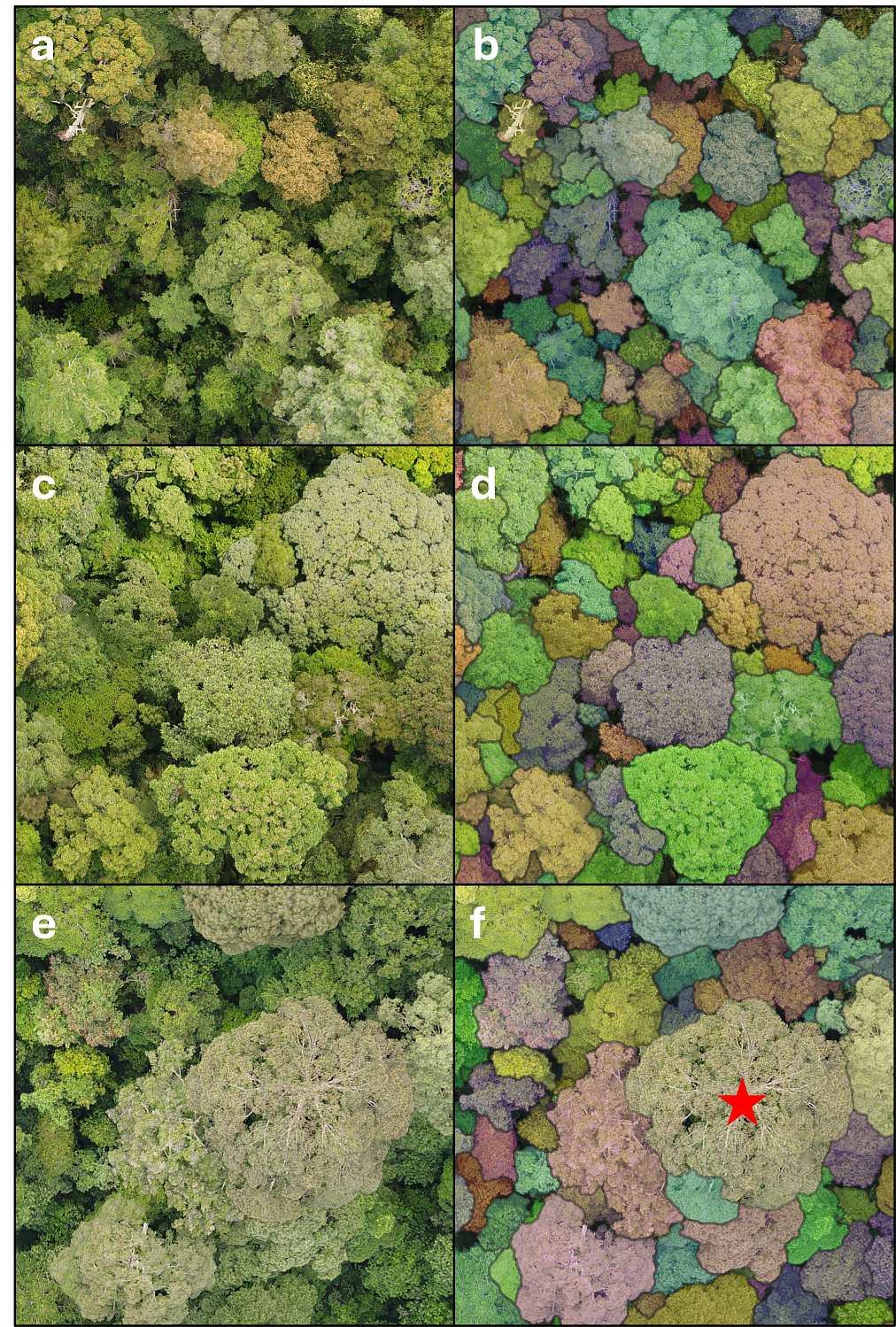}
\caption{Examples of tree crown segmentation results in Borneo Forest. (a, c, e) Original orthomosaic images. (b, d, f) Corresponding inference results overlaid on the images shown in (a), (c), and (e), respectively. Star symbol indicates a tree crown that was not detected by the model.}
\label{fig:borneo_result}
\end{figure}

The developed crown segmentation model showed high generalization performance across various forest types even in geographically distinct regions. We present representative inference results from three selected locations at each inference site (Figures~\ref{fig:sanpoku_result}--\ref{fig:borneo_result}).

In Sanpoku Forest, which consists of temperate deciduous broadleaf forest and conifer plantation, the model achieved high delineation performance (Figure~\ref{fig:sanpoku_result}). The results demonstrate that tree crowns of various sizes are successfully detected. Some dead trees are not individually detected and are instead merged with surrounding trees into a single polygon. In addition, Figure~\ref{fig:sanpoku_result}(e) represents a conifer plantation, where the model also successfully detects individual trees.

Similarly, in Sadayama Forest, which consists of temperate evergreen broadleaf forest, the results demonstrate that tree crowns of various sizes are successfully detected along with the edge of the crowns (Figure~\ref{fig:sadayama_result}).

In Borneo Forest, which consists of lowland dipterocarp forest, the model also showed strong performance, with small trees located between large crowns successfully detected (Figure~\ref{fig:borneo_result}). However, some large crowns were not successfully delineated (Figure~\ref{fig:borneo_result}(e, f)), suggesting limitations of the model.

\vspace{5mm}
\vspace{5mm}

\begin{table}[htbp]
\centering
\caption{Results on the test dataset represent the mean performance across three independent runs with different random seeds. Values in parentheses indicate the standard deviation (SD) of these runs. All metrics are multiplied by 100.}
\label{tab:model_performance}

\begin{tabular}{lccc}
\toprule
Backbone & mAP & mAP50 & mAP75 \\
\midrule

ResNet-50 &
17.8 ($\pm$1.1) &
42.3 ($\pm$1.0) &
13.1 ($\pm$1.3) \\

ResNet-101 &
18.0 ($\pm$2.2) &
42.1 ($\pm$3.4) &
12.7 ($\pm$3.3) \\

Swin Transformer-Tiny &
18.7 ($\pm$1.9) &
44.3 ($\pm$3.4) &
12.9 ($\pm$1.3) \\

Swin Transformer-Small &
23.1 ($\pm$0.8) &
49.6 ($\pm$0.7) &
19.0 ($\pm$1.7) \\

\bottomrule
\end{tabular}
\end{table}

\clearpage

\clearpage
\section{Discussion}

Our developed model showed promising crown segmentation performance across various forest types in geographically and biologically distinct regions, even though it was trained on a limited number of study sites. In particular, the high generalizability, the detailed polygon generation reflecting crown shapes, and the minimal omission of small trees can be attributed to the high quality and quantity of the training data.

In Japan, natural and plantation forests account for approximately 60\% and 40\% of the forest area, respectively \cite{ForestryAgency2024}. Natural forests include subarctic evergreen conifer forests, temperate evergreen and deciduous broadleaf forests, and subtropical evergreen forests, among which temperate forests occupy the largest area \cite{ForestryAgency2024}. In plantation forests, approximately 70\% consists of coniferous species such as Japanese cedar and cypress \cite{ForestryAgency2024}. Considering that the orthomosaic images from Sanpoku Forest and Sadayama Forest include natural deciduous and evergreen broadleaf forests, as well as plantations of evergreen conifers, the developed model is expected to provide practical segmentation performance across most forest areas in Japan.

Since our primary objective was to develop a crown segmentation model for broadleaf forests, the training dataset was highly biased toward broadleaf-dominated forests. Nevertheless, the segmentation performance for coniferous species was also high. This result suggests that the structural diversity captured in our broadleaf-dominated training dataset may be sufficient to represent the crown morphological characteristics of conifers. The inclusion of a substantial number of conifer crowns in the training dataset likely further enhances segmentation performance in coniferous stands.

The developed model showed high segmentation performance on the orthomosaic image from Borneo Forest, although the training dataset included only images from Japanese forests. This result suggests that a model trained on data from an ecologically diverse region such as Japan can be applied to other regions with comparable performance. However, inference accuracy was lower for crowns with large crown diameter, or white-colored crowns, which were not included in the training dataset from Japan (Figure~\ref{fig:borneo_result}(e, f)). This limitation could be mitigated in future work by incorporating more diverse broadleaf forest data from regions worldwide into the training dataset.

This study demonstrated that practical crown segmentation performance for broadleaf trees can be achieved using only three RGB channels, without explicitly incorporating canopy structure information such as canopy height model (CHM). The result is consistent with a previous study showing only marginal improvement in tree crown segmentation performance in plantations in Canada when structural information was incorporated as a fourth channel in a Mask R-CNN \cite{Teng2025}. The characteristics of CHMs can vary depending on the source data, sensor type, and processing workflow, resulting in differences in canopy-height accuracy, surface smoothness, and noise levels. In addition, spatial misalignment between CHM and RGB imagery occasionally occurred, which can increase the risk of introducing noise into the training dataset. Considering these concerns, we propose using CHMs as supplementary references during annotations or in post-processing step after inference, rather than as input data for the model, due to variability in data quality across datasets.

In future work, we plan to incorporate sufficient dataset on coniferous forests, in addition to broadleaf forests, aiming to develop a more generalizable model applicable to a broader range of forest types across the world. In particular, expanding the dataset for coniferous forests is important, given their primary role in forestry in Japan and their broad distribution in high-latitude forests globally. Overall, the establishment of high-quality and precisely annotated training datasets will be key to achieving high segmentation performance.

\section{Conclusion}

In this study, we developed an instance segmentation model for individual tree crowns in broadleaf forests and evaluated its generalization performance. We constructed a high-quality annotated dataset comprising 18,507 crown polygons across multiple forest types in Japan, and developed a model based on the Mask2Former architecture. The developed model achieved high segmentation performance even in structurally complex broadleaf forests, as well as in geographically and biologically distinct regions. These findings demonstrate that the development of large, high-quality, and diverse training datasets is essential for building high-performance segmentation models.

In future work, we aim to further improve performance and expand its applicability by incorporating more diverse forest-type data into the training dataset. The model has been implemented in the forest analysis software ``DF Scanner Pro,'' supporting practical forest management using UAVs.

\section*{Acknowledgements}

We are grateful to Sabah Forestry Department and Sabah Forest Research Centre for their generous support in conducting this study. We further thank Rikuto Ueno, Shiori Niitsuma, Midori Watanabe, and Naoto Shinohara for their valuable contributions as annotators.

\section*{Funding Sources}

This work was supported by Kyoto Prefecture, the City of Kyoto, KYOTO Industrial Support Organization 21, Kyoto Wisdom Industry Creation Center, and the Advanced Science, Technology \& Management Research Institute of KYOTO. This work was also supported by the Small Business Innovation Research (SBIR) Program (Project Number: JPJ010717), administered by the National Agriculture and Food Research Organization and NEDO.

This work was partly supported by Japan Society for the Promotion of Science KAKENHI (Grant Numbers: 21H05314 and 21H02564) to Yusuke Onoda. This study was also supported by the Grant for Global Sustainability (GGS) from Institute for the Advanced Study of Sustainability, United Nations University (UNU-IAS), to Kanehiro Kitayama.

\section*{Author Contributions}

Conceptualization: M.N., M.O.; Training data preparation: M.N., T.I.; AI model training: M.N., Y.M.; Support for data acquisition: Y.O., R.T., K.K.H., K.K., R.O.; Software implementation: K.I.; Writing -- original draft: M.N.; Writing -- review \& editing: M.N., R.T., M.O.; Supervision: M.O.

All authors have read and agreed to the published version of the manuscript.

%Bibliography
\bibliographystyle{unsrt}  
\bibliography{references}

@article{Ball2023,
  title={Accurate delineation of individual tree crowns in tropical forests from aerial RGB imagery using Mask R-CNN},
  author={Ball, James G.C. and Hickman, Sebastian H.M. and Jackson, Tobias D. and Koay, Xian Jing and Hirst, James and Jay, William and Archer, Matthew and Aubry-Kientz, Mélaine and Vincent, Grégoire and Coomes, David A.},
  journal={Remote Sensing in Ecology and Conservation},
  volume={9},
  number={5},
  pages={641--655},
  year={2023},
  doi={10.1002/rse2.332}
}

@article{Brieger2019,
  title={Advances in the derivation of Northeast Siberian forest metrics using high-resolution UAV-based photogrammetric point clouds},
  author={Brieger, Frederic and Herzschuh, Ulrike and Pestryakova, Luidmila A. and Bookhagen, Bodo and Zakharov, Evgenii S. and Kruse, Stefan},
  journal={Remote Sensing},
  volume={11},
  number={12},
  year={2019},
  doi={10.3390/rs11121447}
}

@article{Chen2021,
  title={Geomorphological Analysis Using Unpiloted Aircraft Systems, Structure from Motion, and Deep Learning},
  author={Chen, Zhiang and Scott, Tyler R. and Bearman, Sarah and Anand, Harish and Keating, Devin and Scott, Chelsea and Arrowsmith, J Ramon and Das, Jnaneshwar},
  journal={arXiv},
  year={2021},
  doi={10.1109/IROS45743.2020.9341354}
}

@article{Cheng2022,
  title={Masked-attention Mask Transformer for Universal Image Segmentation},
  author={Cheng, Bowen and Misra, Ishan and Schwing, Alexander G. and Kirillov, Alexander and Girdhar, Rohit},
  journal={arXiv},
  year={2022}
}

@article{Dai2018,
  title={A new method for 3D individual tree extraction using multispectral airborne LiDAR point clouds},
  author={Dai, Wenxia and Yang, Bisheng and Dong, Zhen and Shaker, Ahmed},
  journal={ISPRS Journal of Photogrammetry and Remote Sensing},
  volume={144},
  pages={400--411},
  year={2018},
  doi={10.1016/j.isprsjprs.2018.08.010}
}

@article{Do2026,
  title={A UAV RGB dataset and method for instance tree crown segmentation for biodiversity monitoring},
  author={Do, Mai Viet Hoang and Phung, Duc-Thang and Pham, Hoang Duy Linh and Pham, Quang-Duy and Hoang, Van-Nam and Hoang, Van-Sam and Vlaminck, Michiel and Luong, Hiep and Tran, Thanh-Hai and Vu, Hai and Le, Thi-Lan},
  journal={Scientific Reports},
  year={2026},
  doi={10.1038/s41598-026-36541-y}
}

@techreport{ForestryAgency2024,
  title={Annual report on forest and forestry in Japan fiscal year 2024 (Summary)},
  author={Forestry Agency},
  institution={Ministry of Agriculture, Forestry and Fisheries, Japan},
  year={2024}
}

@article{He2015,
  title={Deep Residual Learning for Image Recognition},
  author={He, Kaiming and Zhang, Xiangyu and Ren, Shaoqing and Sun, Jian},
  journal={arXiv},
  year={2015}
}

@article{Htoo2025,
  title={Development of crown-based allometric equations for estimating stem diameter and above-ground biomass using UAV-LiDAR in 23 species-rich natural forests of Japan},
  author={Htoo, Kyaw Kyaw and Onishi, Masanori and Rahman, Md Farhadur and Takeshige, Ryuichi and Kitajima, Kaoru and Onoda, Yusuke},
  journal={Journal of Forest Research},
  volume={30},
  number={6},
  pages={491--501},
  year={2025},
  doi={10.1080/13416979.2025.2576384}
}

@article{Koch2006,
  title={Detection of Individual Tree Crowns in Airborne Lidar Data},
  author={Koch, Barbara and Heyder, Ursula and Weinacker, Holger},
  journal={Photogrammetric Engineering \& Remote Sensing},
  year={2006}
}

@article{Lin2015,
  title={Microsoft COCO: Common Objects in Context},
  author={Lin, Tsung-Yi and Maire, Michael and Belongie, Serge and Bourdev, Lubomir and Girshick, Ross and Hays, James and Perona, Pietro and Ramanan, Deva and Zitnick, C. Lawrence and Dollár, Piotr},
  journal={arXiv},
  year={2015}
}

@article{Liu2021,
  title={Swin Transformer: Hierarchical Vision Transformer using Shifted Windows},
  author={Liu, Ze and Lin, Yutong and Cao, Yue and Hu, Han and Wei, Yixuan and Zhang, Zheng and Lin, Stephen and Guo, Baining},
  journal={arXiv},
  year={2021}
}

@article{Loshchilov2019,
  title={Decoupled Weight Decay Regularization},
  author={Loshchilov, Ilya and Hutter, Frank},
  journal={arXiv},
  year={2019},
  doi={10.48550/arXiv.1711.05101}
}

@article{Nevalainen2017,
  title={Individual tree detection and classification with UAV-Based photogrammetric point clouds and hyperspectral imaging},
  author={Nevalainen, Olli and Honkavaara, Eija and Tuominen, Sakari and Viljanen, Niko and Hakala, Teemu and Yu, Xiaowei and Hyyppä, Juha and Saari, Heikki and Pölönen, Ilkka and Imai, Nilton N. and Tommaselli, Antonio M.G.},
  journal={Remote Sensing},
  volume={9},
  number={3},
  year={2017},
  doi={10.3390/rs9030185}
}

@article{Onishi2018,
  title={Automatic classification of trees using a UAV onboard camera and deep learning},
  author={Onishi, Masanori and Ise, Takeshi},
  journal={arXiv},
  year={2018}
}

@article{Payer2019,
  title={Segmenting and tracking cell instances with cosine embeddings and recurrent hourglass networks},
  author={Payer, Christian and Štern, Darko and Feiner, Marlies and Bischof, Horst and Urschler, Martin},
  journal={Medical Image Analysis},
  volume={57},
  pages={106--119},
  year={2019},
  doi={10.1016/j.media.2019.06.015}
}

@article{Steier2025,
  title={Comparison of manual and semi-automated synthetic training data creation for individual tree crown delineation},
  author={Steier, Janik and Iwaszczuk, Dorota},
  journal={The International Archives of the Photogrammetry, Remote Sensing and Spatial Information Sciences},
  volume={XLVIII-1/W6-2025},
  pages={227--233},
  year={2025},
  doi={10.5194/isprs-archives-xlviii-1-w6-2025-227-2025}
}

@article{Takeshige2025,
  title={High-resolution digital canopy height models, terrain models, ortho-mosaic photos, and canopy tree crown shapes derived from UAV-borne LiDAR at 22 tree census plots across Japanese natural forests},
  author={Takeshige, Ryuichi and Htoo, Kyaw Kyaw and Onishi, Masanori and Rahman, Farhadur Md and Hoshizaki, Kazuhiko and Ida, Hideyuki and Ishihara, Masae Iwamoto and Itoh, Akira and Kaneko, Takayuki and Katayama, Ayumi and Kuramoto, Shigeo and Kurokawa, Hiroko and Maki, Masayuki and Masaka, Kazuhiko and Nakaji, Tatsuro and Nakamura, Masahiro and Nishimura, Naoyuki and Noguchi, Mahoko and Sakai, Atsushi and Takashima, Atsushi and Tashiro, Naoaki and Tokuchi, Naoko and Yamagawa, Hiromi and Onoda, Yusuke},
  journal={Ecological Research},
  volume={40},
  number={4},
  pages={657--670},
  year={2025},
  doi={10.1111/1440-1703.12555}
}

@article{Teng2025,
  title={Assessing SAM for Tree Crown Instance Segmentation from Drone Imagery},
  author={Teng, Mélisande and Ouaknine, Arthur and Laliberté, Etienne and Bengio, Yoshua and Rolnick, David and Larochelle, Hugo},
  journal={arXiv},
  year={2025}
}

@article{Weinstein2020,
  title={Cross-site learning in deep learning RGB tree crown detection},
  author={Weinstein, Ben G. and Marconi, Sergio and Bohlman, Stephanie A. and Zare, Alina and White, Ethan P.},
  journal={Ecological Informatics},
  volume={56},
  year={2020},
  doi={10.1016/j.ecoinf.2020.101061}
}

@article{Yang2020,
  title={An Individual Tree Segmentation Method Based on Watershed Algorithm and Three-Dimensional Spatial Distribution Analysis from Airborne LiDAR Point Clouds},
  author={Yang, Juntao and Kang, Zhizhong and Cheng, Sai and Yang, Zhou and Akwensi, Perpetual Hope},
  journal={IEEE Journal of Selected Topics in Applied Earth Observations and Remote Sensing},
  pages={1055--1067},
  year={2020},
  doi={10.1109/JSTARS.2020.2979369}
}

@article{Zhen2015,
  title={Agent-based region growing for individual tree crown delineation from airborne laser scanning (ALS) data},
  author={Zhen, Zhen and Quackenbush, Lindi J. and Stehman, Stephen V. and Zhang, Lianjun},
  journal={International Journal of Remote Sensing},
  volume={36},
  number={7},
  pages={1965--1993},
  year={2015},
  doi={10.1080/01431161.2015.1030043}
}

@article{Zheng2025,
  title={A Review of Individual Tree Crown Detection and Delineation From Optical Remote Sensing Images: Current progress and future},
  author={Zheng, Juepeng and Yuan, Shuai and Li, Weijia and Fu, Haohuan and Yu, Le and Huang, Jianxi},
  journal={IEEE Geoscience and Remote Sensing Magazine},
  volume={13},
  number={1},
  pages={209--236},
  year={2025},
  doi={10.1109/MGRS.2024.3479871}
}

@article{Zhou2020,
  title={Individual tree parameters estimation for plantation forests based on UAV oblique photography},
  author={Zhou, Xuemei and Zhang, Xiaoli},
  journal={IEEE Access},
  volume={8},
  pages={96184--96198},
  year={2020},
  doi={10.1109/ACCESS.2020.2994911}
}

\end{document}